# Nanoelectromechanical Systems (NEMS) for Hardware Security in Advanced Packaging

Himanandhan Reddy Kottur*, Pavanbabu Arjunamahanthi*, M. Shafkat M. Khan*, Liton Kumar Biswas*, Nitin Varshney* and Navid Asadizanjani*
*Department of Electrical and Computer Engineering, University of Florida
5000 Malachowsky Hall, 1889 Museum Road Gainesville, Florida-32611, USA
Contact Author Email: h.kottur@ufl.edu

## Abstract

As hardware security threats escalate across semiconductor manufacturing and advanced packaging, there is a growing need for novel physical mechanisms to counter sophisticated attacks such as tampering, counterfeiting, and supply chain infiltration. This paper presents Nanoelectromechanical Systems (NEMS) as an emerging class of hardware security primitives that enable physical assurance, tamper detection, and authentication at the device level. Leveraging mechanisms such as NEMS-based Physically Unclonable Functions (PUFs), shape memory materials, resonance-based fingerprints, and physical unlocking architectures, these systems offer enhanced resilience to reverse engineering, side-channel attacks, and environmental degradation. By harnessing mechanical unpredictability and fabrication-induced nanoscale variability, NEMS technologies introduce a physically robust and low-power alternative to conventional digital security methods. Their seamless integration into standard semiconductor workflows paves the way for scalable, verifiable, and secure solutions across defense, aerospace, critical infrastructure, and consumer electronics.

## I. Introduction

The rapid advancement of semiconductor technology and the increasing complexity of electronic systems have elevated the demand for robust, hardware-level security solutions [1–4]. Traditional mechanisms, though effective to some extent, are increasingly susceptible to reverse engineering, side-channel attacks, and physical tampering [5,6]. As these threats evolve, nanoelectromechanical systems (NEMS) have emerged as a promising technology to address critical challenges in semiconductor security and advanced packaging [7–9]. NEMS devices, categorized with characteristic length of less than 100 nm, operate at the nanometer scale, integrating electrical and mechanical functionalities for sensing, actuation, and signal processing. Their inherent characteristics—including high sensitivity to physical perturbations, favorable surface-area-to-volume ratios, and CMOS compatibility—make them suitable for hardware security applications.

Specifically, NEMS enables device authentication, tamper detection, and supply chain assurance. Process-induced variations during fabrication allow NEMS to generate physically unclonable functions (PUFs) that offer high entropy and resistance to duplication [11, 12]. Hwang et al. first introduced the concept of NEMS-PUFs, leveraging stiction effects caused by van der Waals forces in suspended nanowires positioned between contact gates. These physical effects, often problematic in MEMS/NEMS design, are repurposed here as entropy sources for secure identification [10].

In packaging environments, the mechanical responsiveness of NEMS to pressure, vibration, and electromagnetic interference enables real-time, passive tamper detection [13]. This capability adds an embedded, hardware-rooted layer of physical assurance to microelectronic systems. The global OSAT (Outsourced Semiconductor Assembly and Test) industry, which handles a significant portion of advanced packaging, continues to see multibillion-dollar growth, as shown in Figure 1. Integrating NEMS into these workflows positions them not only as sensors or actuators but also as key enablers of secure electronics.

This paper explores the integration of NEMS in advanced packaging, focusing on their use as security primitives. It reviews their operating principles, presents specific security applications, and highlights recent developments supporting their feasibility. Given the limitations of current primitives—low entropy, high power consumption, and ease of reverse engineering—this work proposes mechanically rooted NEMS-based solutions to enhance hardware resilience across sectors such as defense, aerospace, consumer electronics, and healthcare.

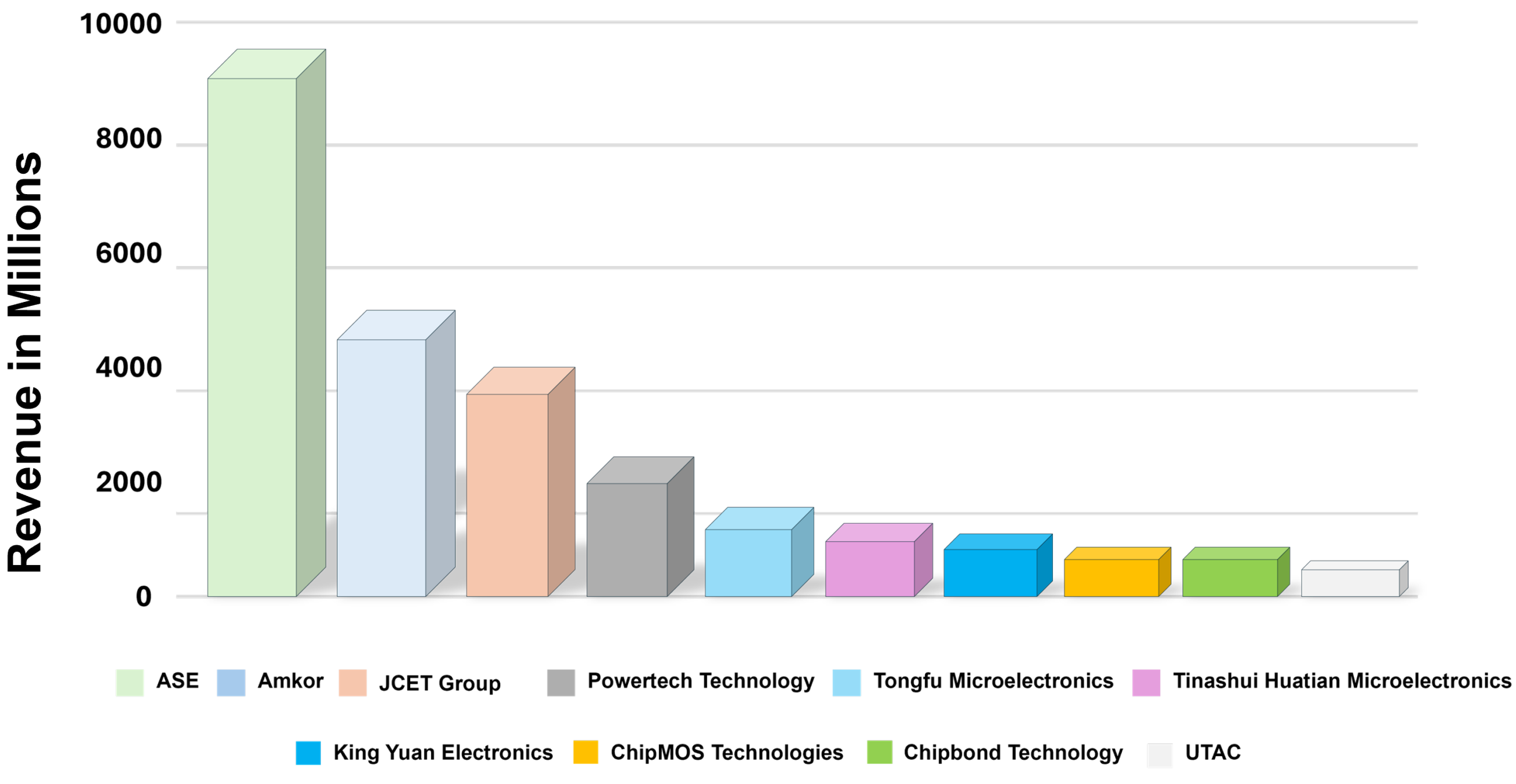

Figure 1: Global revenue (in millions USD) of major advanced packaging OSATs companies [14]

## II. Challenges in Semiconductor Industry

### A. *Counterfeiting*

Counterfeiting remains a persistent threat in the semiconductor ecosystem. Unauthorized entities often manufacture or refurbish components to mimic authentic ones, leading to the infiltration of substandard or malicious chips into critical systems [15]. For example, a 2012 U.S. Senate Armed Services Committee investigation identified over 1,800 cases of suspected counterfeit electronics in military systems, including mission computers for fighter jets and thermal weapon sights [16]. These components compromised system reliability and posed serious risks by enabling potential insertion of backdoors or sabotage features.

### B. *Tampering*

Tampering and unauthorized modifications present another significant concern. Once a chip exits the foundry and enters packaging and distribution, it becomes vulnerable to physical attacks such as microprobing, laser fault injection, focused ion beam (FIB) editing, and other invasive techniques. These can be used to extract secret keys, alter circuit functionality, or insert malicious payloads. A widely cited example is the attack on the PlayStation 3, where fault injection enabled bypassing encryption and gaining unauthorized access to secure content. In industrial or defense applications, similar attacks could lead to IP theft, denial-of-service, or unauthorized control of systems.

### C. *Supply Chain Vulnerabilities*

Supply chain vulnerabilities further exacerbate security risks, as chips often traverse multiple entities—spanning continents—before deployment. Each transition between design, fabrication, packaging, testing, and integration introduces potential attack vectors [17]. A notable case, the 2018 Bloomberg report (though later disputed), alleged hardware backdoors embedded in Supermicro server boards used by Amazon and Apple [18]. While the case remains controversial, it underscored widespread concerns regarding supply chain infiltration, a threat substantiated by other documented incidents [19].

Traditional hardware security approaches, such as cryptographic authentication and electronic PUFs (ePUFs), offer limited protection in such environments. Cryptographic methods rely on securely stored keys, which are vulnerable to side-channel attacks, including power analysis, timing, and electromagnetic leakage. High-profile vulnerabilities like Spectre and Meltdown exposed the susceptibility of even advanced processor architectures to such attacks.

Moreover, ePUFs—based on variations in electronic properties like delay paths or threshold voltages—are sensitive to aging, temperature, and radiation. These effects degrade their reliability over time, especially in high-stress environments such as aerospace and defense, where stable key storage and authentication are critical.

In contrast, NEMS-based security primitives provide a fundamentally different paradigm. Their nanoscale mechanical structures are inherently resistant to traditional electronic attack vectors and offer unique, unpredictable responses due to fabrication-induced variability. Properties such as resonance frequency, deflection profiles, and structural deformation can serve as robust physical identifiers.

Crucially, these characteristics are physically responsive—any invasive tampering (e.g., FIB-based probing) can subtly alter the mechanical structure, changing its unique signature and thus providing tamper evidence. Additionally, NEMS devices are compatible with standard CMOS processes, enabling seamless integration into conventional packaging and back-end flows. It makes them scalable, energy-efficient, and suitable for deployment in high-reliability and security-critical environments.

## III. NEMS-Based PUFs

One of the most promising applications of NEMS in hardware security is the development of NEMS-based physically unclonable functions (PUFs). Unlike conventional electronic PUFs that rely on threshold voltage variations or gate delays, NEMS-PUFs utilize intrinsic mechanical and structural irregularities introduced during nanoscale fabrication. These include differences in material composition, surface roughness, grain boundaries, and mechanical stiffness [20]. Such variations create unique, unclonable mechanical fingerprints that persist even under identical manufacturing conditions.

Because their identity is rooted in physical structure rather than electrical behavior, NEMS-PUFs offer inherent resistance to invasive and side-channel attacks. Physical probing methods—such as focused ion beam (FIB) editing or microprobing—are likely to alter or destroy the delicate mechanical features that encode their identity. It makes NEMS-PUFs intrinsically tamper-evident and robust against physical intrusion. Additionally, unlike ePUFs that can degrade due to thermal or electrical stress, the structural integrity of NEMS devices remains more stable under environmental variations, enhancing long-term reliability.

A further advantage of NEMS-PUFs is their ability to generate high entropy. The multidimensional nature of nanoscale mechanical variations results in highly unpredictable responses, ideal for cryptographic key generation. These physically derived keys are challenging to model, duplicate, or brute-force, making them well-suited for secure key storage, device identification, and access control in sensitive systems [21].

NEMS-based locking architectures also enable chiplet-level access control in 2.5D and 3D heterogeneous integration platforms. These mechanical structures can be embedded within the die, interposer, or substrate layers, assigning each chip package a unique physical signature. It strengthens authentication and mitigates supply chain threats, including unauthorized part substitution and counterfeiting. In defense electronics and critical infrastructure, NEMS-PUFs provide a hardware-level verification method that does not rely on external memory or software-based checks.

The benefits of NEMS-PUFs extend to multiple sectors. In defense, they can secure field-deployed systems through immutable hardware authentication. In finance, they can protect secure modules like smart cards and payment terminals from cloning or tampering. Industrial control systems can use NEMS-PUFs to anchor trust at the hardware level, reducing risks of firmware injection or counterfeit components in critical operations. Overall, the unique physical nature of NEMS-PUFs makes them highly effective for modern hardware security. By leveraging unavoidable manufacturing randomness, they create tamper-resistant, unclonable device identities. Their seamless integration into semiconductor packaging provides a scalable and manufacturable foundation for embedding trust directly into microelectronic systems.

## IV. Shape Memory Materials for Tamper Detection and Authentication

An emerging direction in hardware security involves integrating shape memory materials into NEMS-based architectures. Shape Memory Alloys (SMAs) and Shape Memory Polymers (SMPs) possess the ability to undergo reversible or irreversible transformations in shape or mechanical state in response to external stimuli such as temperature, mechanical stress, or electrical input [22]. These unique properties enable novel tamper-detection and authentication mechanisms that are both physically embedded and highly sensitive.

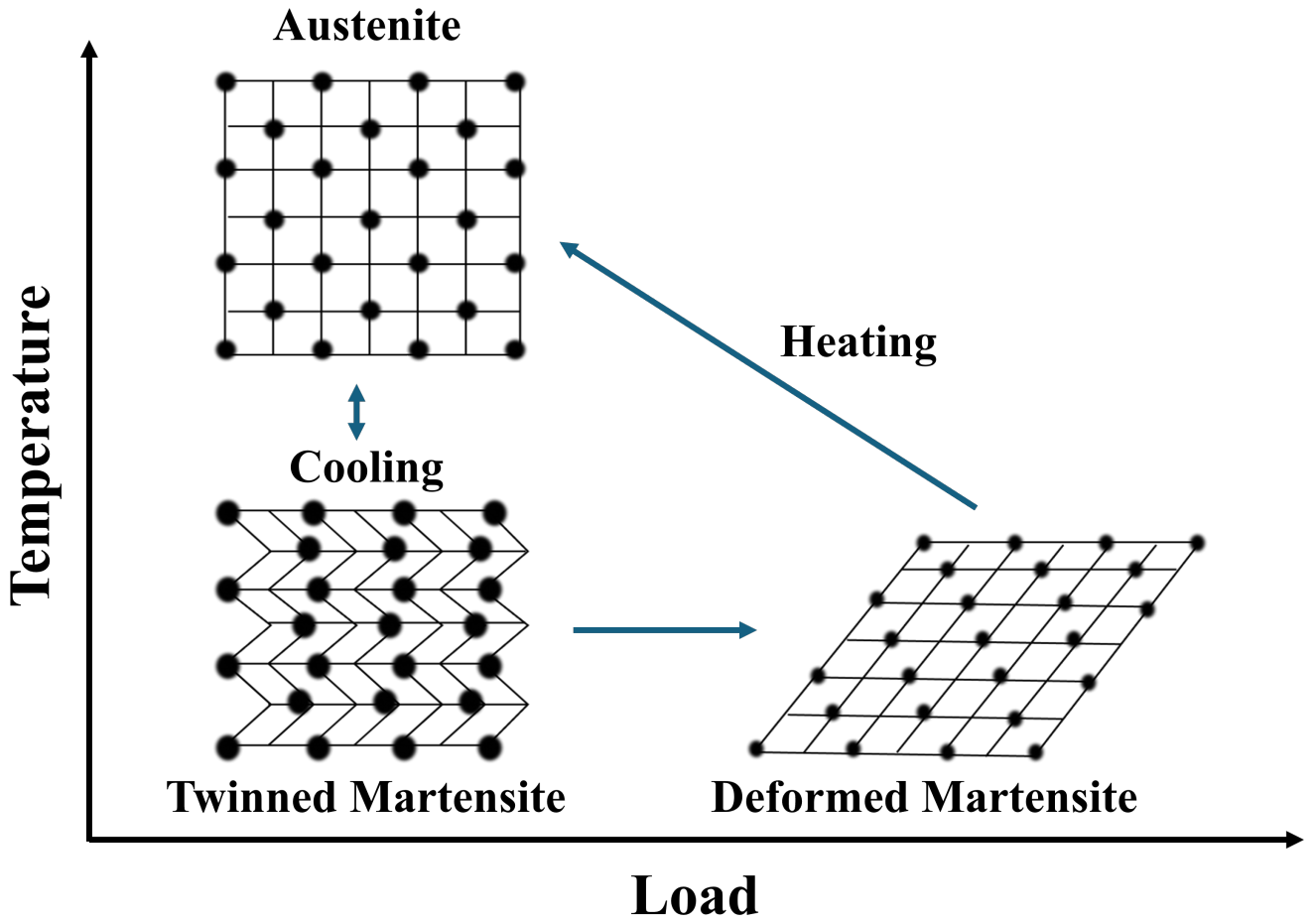

Figure 2: Microstructural views of martensite and austenite phases in SMAs used in packaging. Their reversible transformation enables built-in, tamper-evident security in semiconductor hardware.

Incorporating shape memory materials into security primitives allows for structures that change configuration upon expo-sure to tampering or abnormal conditions. SMAs, for instance, can be engineered to remain in a preset shape under regular operation but permanently shift to a different configuration once a critical threshold—such as a thermal spike or directed heat—is exceeded [23]. This irreversible change acts as a passive, tamper-evident mechanism, leaving behind a permanent physical signature of compromise. Unlike digital logs or software-based detection, these responses are non-resettable and require no power to function, making them well-suited for long-term deployment in field environments.

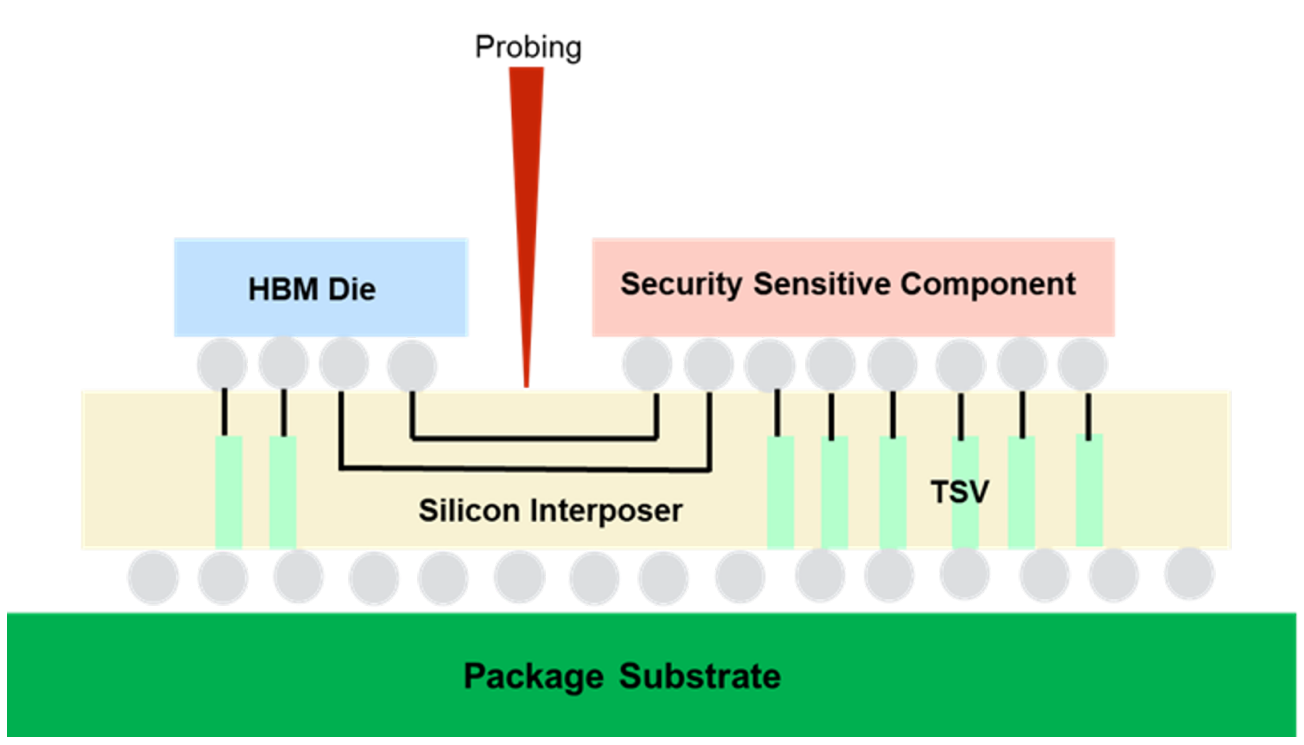

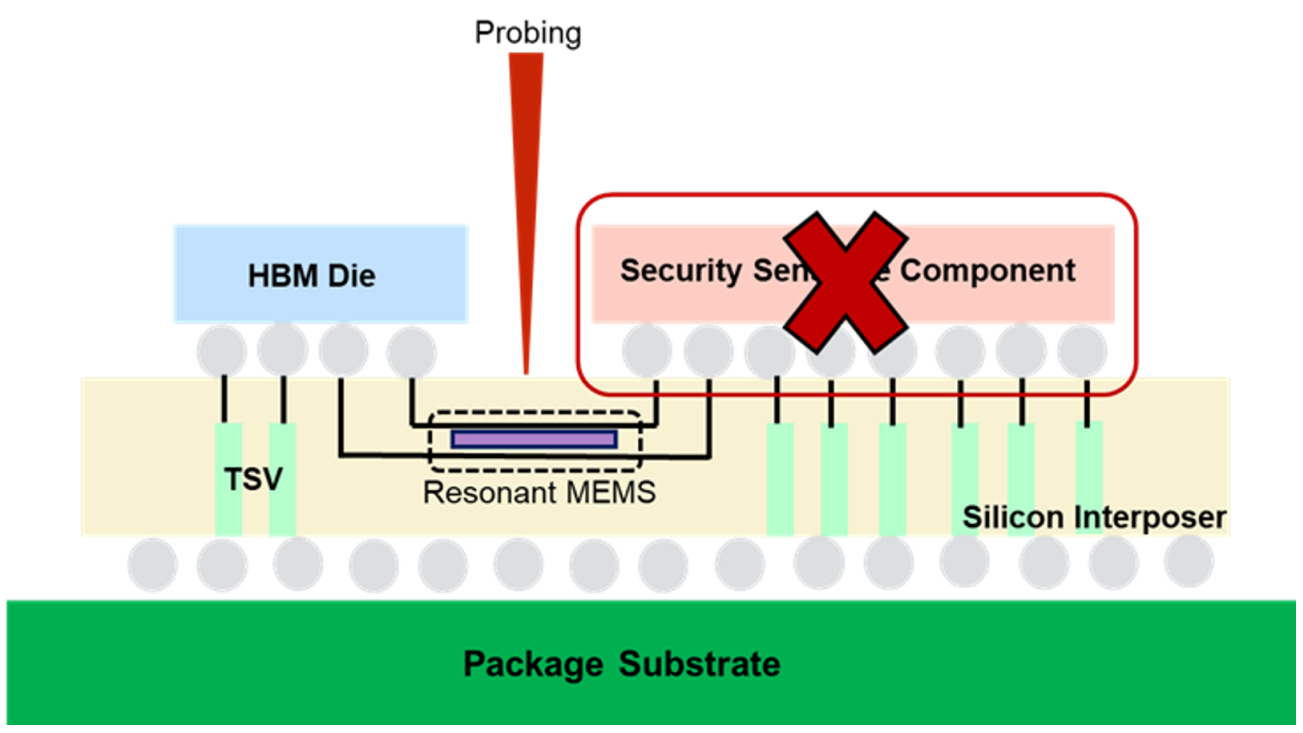

Figure 3: Conventional circuit prone to Microprobing (left) versus NEMS-resonator-enhanced design offering mechanical fingerprinting and tamper resilience (right)

Shape Memory Polymers, as illustrated in Figure 2, offer additional flexibility for tamper detection. SMPs can be designed to deform, expand, or contract when triggered by environmental conditions such as heat or mechanical force. This deformation may serve as a visual indicator or introduce physical misalignment that disrupts access to critical circuit regions. For example, an SMP-based seal could block access to sensitive circuitry upon detecting tampering, rendering the system inoperable or requiring forensic validation before reuse. These passive responses are especially beneficial in untrusted supply chain environments where real-time monitoring is not feasible. Embedding shape memory materials directly into packaging or substrate layers enhances physical security in a persistent, power-independent manner. Unlike software-based methods or active sensing, these materials provide tamper evidence even in off-grid or dormant states. It makes them particularly valuable in defense, aerospace, and secure financial systems where hardware may remain deployed for extended periods under harsh conditions [24]. Ultimately, integrating shape memory materials into NEMS-based security frameworks represents a convergence of smart materials and hardware assurance. These materials enable low-power, physically verifiable, and irreversible responses to tampering, reinforcing the trustworthiness of semiconductor devices. As hardware attacks grow more sophisticated, such embedded physical responses will become increasingly essential in defending against intrusion and ensuring long-term system integrity.

## V. Resonance-Based Fingerprints for Secure Identification

NEMS resonators offer a robust method for hardware authentication by leveraging unique resonance frequency signatures. Each resonator, fabricated through nanoscale processes, inherently contains subtle imperfections arising from uncontrollable variations during manufacturing. These differences result in distinct resonance frequencies and mode shapes, effectively creating a physical fingerprint—analogous to human biometrics such as fingerprints or iris patterns [25, 26]. The uniqueness and non-replicability of these signatures make resonance-based authentication highly secure and extremely difficult to clone or reverse-engineer.

A key advantage of resonance-based authentication lies in its intrinsic resistance to duplication. Figure 3 illustrates how a NEMS-based design resists probing attacks compared to conventional structures. Because the resonance characteristics depend on nanoscale variations in geometry, material composition, and internal stress, even advanced fabrication techniques cannot produce two devices with identical responses. It forms a significant deterrent against counterfeiting and spoofing. Furthermore, mechanical identifiers embedded directly into the packaging or die eliminate the need for additional security ICs or external modules. It reduces system complexity and cost while physically anchoring the identity to the device itself.

Another significant benefit is the extremely low power consumption. Unlike traditional cryptographic systems that require active power to maintain keys or state, resonance-based methods operate using passive mechanical properties. The resonator can be excited and read with minimal energy, making the approach ideal for low-power or energy-harvesting applications such as IoT devices and wearables [27].

When integrated with other NEMS-based technologies—such as NEMS-PUFs and shape memory materials—resonance-based fingerprints enhance the security stack by introducing a complementary, physically derived identifier. This multilayered approach provides dynamic tamper resistance, secure authentication, and enduring identity across

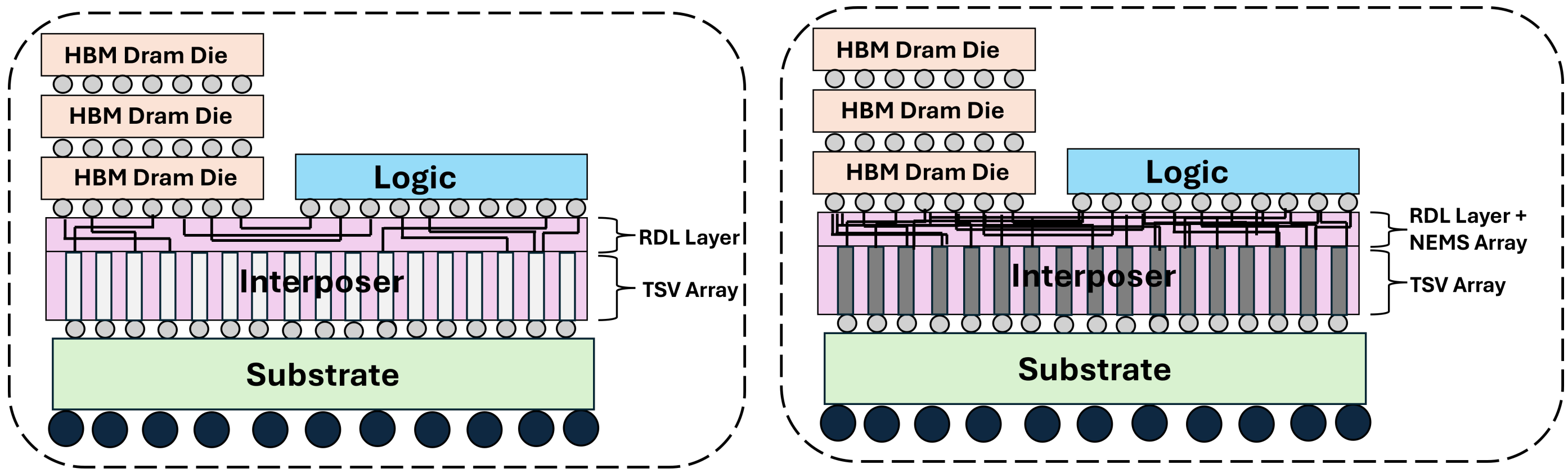

Figure 4: Illustration of locking/unlocking mechanisms using NEMS integrated into the redistribution layer (RDL) [28]. These physical-layer triggers enable secure, condition-based activation of chip functionality

the device lifecycle. From fabrication and supply chain logistics to final deployment, this strategy strengthens hardware trust and reliability in mission-critical sectors.

## VI. NEMS-Enabled Locking and Unlocking Mechanisms

One of the most innovative security applications of NEMS lies in the development of sophisticated locking and unlocking mechanisms embedded within semiconductor devices, as illustrated in Figure 4. These mechanisms are designed to respond selectively to external triggers—such as specific temperature thresholds, mechanical forces, or electrical signals—allowing controlled activation or deactivation of particular chip functionalities. For example, a semiconductor device could remain in a locked or dormant state until it is exposed to a precisely defined thermal or electrical stimulus, ensuring that only authorized personnel or systems can unlock and access its full operational capabilities. This approach effectively prevents unauthorized use or tampering by introducing a physical barrier to activation that cannot be bypassed through conventional electronic hacking methods.

The potential applications of such NEMS-enabled locking mechanisms span critical areas of hardware security. In secure boot processes, they can prevent unauthorized code execution during system startup by ensuring that the device only boots when properly unlocked, mitigating risks posed by malware injection or firmware tampering. Similarly, encryption modules can leverage these mechanisms to keep cryptographic keys physically inaccessible unless the device has been unlocked through authorized means, adding a robust hardware-level safeguard to cryptographic security. Furthermore, NEMS-based locking techniques provide valuable protection against reverse engineering by preventing the unauthorized analysis or replication of sensitive hardware designs—a critical safeguard for intellectual property in competitive industries. By integrating these advanced locking features, semiconductor manufacturers gain the ability to enforce stronger access control, enhancing overall device security and resilience against increasingly sophisticated cyber-physical attacks. These architectures enable hardware-level access control, enhancing physical assurance within semiconductor packaging.

## VII. Integration and Reliability Challenges of NEMS in Advanced Packaging

Integrating NEMS devices into advanced packaging introduces a unique set of engineering challenges that demand precise control over every stage of the process. These devices operate at sub-100 nm dimensions and exhibit extreme sensitivity to mechanical, thermal, and chemical disturbances, making them highly vulnerable during conventional packaging steps such as wafer thinning, die attach, underfill dispensing, molding, and solder reflow. Even minor particulate contamination, sub-micrometer misalignment, or uncontrolled thermal cycling can induce stiction, shift resonance frequencies, or cause irreversible structural damage.

Addressing these vulnerabilities requires process adaptations beyond standard OSAT workflows. Low-damage encapsulation methods, particle-free handling environments, and low-temperature bonding techniques are often necessary to maintain structural and functional integrity. However, such adaptations increase cost, extend manufacturing cycle times, and complicate integration into high-volume production without corresponding advances in process standardization.

Reliability concerns are equally critical. NEMS devices, with their high surface-area-to-volume ratios and movable structural elements, are prone to stiction, wear, and [29]. Packaging-induced stresses—particularly those from the coefficient of thermal expansion (CTE) mismatch between device materials, substrates, and encapsulants—can cause warping, microfracture, or performance [30]. Maintaining long-term hermeticity is essential, as moisture ingress, outgassing, or ionic contamination can lead to corrosion, dielectric charging, and unpredictable shifts in device [31].

Integration at the electrical and mechanical interface presents additional complexity. The scale mismatch between NEMS structures and conventional interconnect geometries makes it challenging to achieve low-resistance, mechanically stable contacts without imposing damaging loads. Furthermore, the extreme sensitivity of NEMS amplifies susceptibility to environmental noise—mechanical vibration, acoustic coupling, and electromagnetic interference from adjacent high-speed or RF circuitry, which can degrade PUF entropy or cause authentication errors [32]. Effective mitigation requires careful layout optimization, electromagnetic shielding, and robust signal conditioning circuitry.

Ultimately, ensuring compatibility with existing semiconductor manufacturing flows remains a decisive hurdle. NEMS-based structures must survive the complete packaging sequence without exceeding thermal budgets, contaminating adjacent CMOS devices, or violating alignment tolerances [33]. Overcoming these challenges will require hybrid integration schemes, mechanically compliant hermetic encapsulants, and packaging architectures purpose-built for nanoscale security primitives. Without such innovations, the scalability and long-term deployment of NEMS-enabled hardware security in mission-critical systems will remain limited despite their significant potential.

## VIII. Conclusion

This work presents a comprehensive framework for leveraging NEMS to address critical hardware security challenges in advanced semiconductor packaging. Through the development and application of NEMS-based PUFs, resonance fingerprinting, and shape-memory-enabled tamper detection, we establish a new class of physical security primitives that extend beyond conventional cryptographic approaches. These NEMS-enabled mechanisms offer inherent resistance to cloning, environmental robustness, and low-power operation—attributes essential for modern electronic systems deployed in untrusted or harsh environments. Furthermore, the introduction of physical-layer locking and unlocking architectures enables precise control over device activation, enhancing resilience against unauthorized use and reverse engineering. As advanced packaging technologies evolve, integrating NEMS directly into substrates, interposers, or die-level structures will provide scalable, trustworthy, and intelligent security solutions. This multidimensional strategy positions NEMS as a foundational technology in the future of hardware assurance, supporting both civilian applications and mission-critical systems across aerospace, defense, and secure communications.

## References

[1] Noor, R., Kottur, H. R., Craig, P. J., Biswas, L. K., Khan, M. S. M., Varshney, N., ... & Asadizanjani, N. (2023). Us microelectronics packaging ecosystem: Challenges and opportunities. arXiv preprint arXiv:2310.11651.

[2] Arjunamahanthi, P., Kottur, H. R., Asadizanjani, N., & Woychik, C. (2025, January). Revitalizing Advanced Packaging in America. In 2025 Pan Pacific Strategic Electronics Symposium (Pan Pacific) (pp. 1-7). IEEE.

[3] Khan, M. S. M., Biswas, L. K., Kottur, H. R., Noor, R., Varshney, N., Hastings, N., & Asadizanjani, N. (2025). Toward standardized vulnerability assessment of advanced packaging against probing attacks. IEEE Design & Test.

[4] Asadizanjani, N., Reddy Kottur, H., & Dalir, H. (2025). Introduction to IC Packaging Development and Assembly Processes. In Introduction to Microelectronics Advanced Packaging Assurance (pp. 1-19). Cham: Springer Nature Switzerland.

[5] Khan, M. Shafkat M., et al. "Inherent Hardware Identifiers: Advancing IC Traceability and Provenance in the Multi-Die Era." Journal of Electronic Testing 41.1 (2025): 85-107.

[6] Kottur, H. R., Biswas, L. K., Varshney, N., Ghosh, S., & Asadizanjani, N. (2025, April). Towards zero defects: machine learning-driven solder bump reliability in semiconductor packaging. In Metrology, Inspection, and Process Control XXXIX (Vol. 13426, pp. 876-881). SPIE.

[7] Kottur, H. R., Khan, A. A., Varshney, N., Biswas, L. K., Dalir, H., & Asadizanjani, N. (2024, August). Enhancing the MEMS Gyroscope Physical Assurance Using Quantum Sensing. In 2024 IEEE Research and Applications of Photonics in Defense Conference (RAPID) (pp. 1-2). IEEE.

[8] Kasiselvanathan, M., & Manikandan, A. (2025). Biometric and bio-inspired approaches for MEMS/NEMS enabled self-powered sensors. In self-powered sensors (pp. 171-185). Academic Press.

[9] Khan, A. A., Sahebkar, K., Xi, C., Tehranipoor, M. M., Need, R. F., & Asadizanjani, N. (2022, May). Security challenges of MEMS devices in HI packaging. In 2022 IEEE 72nd Electronic Components and Technology Conference (ECTC) (pp. 2321-2327). IEEE.

[10] Lyshevski, S. E. (2018). MEMS and NEMS: systems, devices, and structures. CRC press.

[11] Hwang, K. M., Park, J. Y., Bae, H., Lee, S. W., Kim, C. K., Seo, M., ... & Choi, Y. K. (2017). Nano-electromechanical switch based on a physical unclonable function for highly robust and stable performance in harsh environments. ACS nano, 11(12), 12547-12552.

[12] Rassay, S., Ramezani, M., Shomaji, S., Bhunia, S., & Tabrizian, R. (2020). Clandestine nanoelectromechanical tags for identification and authentication. Microsystems & Nanoengineering, 6(1), 103.

[13] Khan, A. A., Sahebkar, K., Need, R. F., Tehranipoor, M. M., & Asadizanjani, N. (2023). Reconfigurable NEMS based advance packaging for anti reverse engineering and counterfeiting. IMAPSource Proceedings, 2022(1), 000258-000263.

[14] Anysilicon. (2019, July 3). TOP 25 OSATs Ranking 2018 - AnySilicon. AnySilicon. https://anysilicon.com/top-25-osats-ranking-2018/

[15] K. Huang, J. M. Carulli and Y. Makris, "Counterfeit electronics: A rising threat in the semiconductor manufacturing industry," 2013 IEEE International Test Conference (ITC), Anaheim, CA, USA, 2013, pp. 1-4, doi: 10.1109/TEST.2013.6651880.

[16] U.S. Senate Committee on Armed Services. (2012). Inquiry into counterfeit electronic parts in the Department of Defense supply chain (Report 122-167). Washington, DC: United States Senate. Retrieved 2014, from http://www.armedservices.senate.gov/imo/media/doc/Counterfeit-Electronic-Parts.pdf

[17] Harrison, J., Asadizanjani, N., & Tehranipoor, M. (2021). On malicious implants in PCBs throughout the supply chain. Integration, 79, 12-22.

[18] Mehta, D., Lu, H., Paradis, O. P., MS, M. A., Rahman, M. T., Iskander, Y., ... & Asadizanjani, N. (2020). The big hack explained: Detection and prevention of PCB supply chain implants. ACM Journal on Emerging Technologies in Computing Systems (JETC), 16(4), 1-25.

[19] Varshney, N., Ghosh, S., Craig, P., Kottur, H. R., Dalir, H., & Asadizanjani, N. (2024). Challenges and opportunities in non-destructive characterization of stacked ic packaging: insights from sam and 3d x-ray analysis. Developments in X-Ray Tomography XV, 13152, 92-99.

[20] Tang, H. Y., Lu, Y., Jiang, X., Ng, E. J., Tsai, J. M., Horsley, D. A., & Boser, B. E. (2016). 3-D ultrasonic fingerprint sensor-on-a-chip. IEEE Journal of Solid-State Circuits, 51(11), 2522-2533.

[21] McGrath, T., Bagci, I. E., Wang, Z. M., Roedig, U., & Young, R. J. (2019). A puf taxonomy. Applied physics reviews, 6(1), 011303.

[22] Moumni, Z., Van Herpen, A., & Riberty, P. (2005). Fatigue analysis of shape memory alloys: energy approach. Smart Materials and Structures, 14(5), S287.

[23] Theiß, R., Lichte, D., & Wolf, K. D. (2012, September). Application of Shape Memory Alloys in Locking Devices. In Smart Materials, Adaptive Structures and Intelligent Systems (Vol. 45103, pp. 425-432). American Society of Mechanical Engineers.

[24] Kim, C., Lee, J. W., Park, G. T., Shin, M. S., Baek, S. H., Woo, J. S., & Choi, W. Y. (2025). In Situ XOR Encryption for Lightweight Security Using Nanoelectromechanical Physically Unclonable Functions. Advanced Intelligent Systems, 2400805.

[25] Khan, M. S. M., Hart, T. E., Kottur, H. R., Biswas, L. K., Varshney, N., Larmann, S. R., & Asadizanjani, N. (2025, May). Thermo-Seal: A Multi-Layered Infrared Watermarking Scheme for on-Field IC Provenance Verification Using Thermal Signatures. In 2025 IEEE 75th Electronic Components and Technology Conference (ECTC) (pp. 2139-2146). IEEE.

[26] Kottur, H. R., Khan, M. S. M., Arjunamahanthi, P., Biswas, L. K., Varshney, N., Hayes, L. S., ... & Asadizanjani, N. (2025, March). MEMS-Enabled Secure PCB Supply Chain. In 2025 IEEE Workshop on Microelectronics and Electron Devices (WMED) (pp. 1-4). IEEE.

[27] Asadizanjani, N., Reddy Kottur, H., & Dalir, H. (2025). Quantum Computing, Wearables, and Next-Gen IC Packaging. In Introduction to Microelectronics Advanced Packaging Assurance (pp. 161-180). Cham: Springer Nature Switzerland.

[28] Asadizanjani, N., Reddy Kottur, H., & Dalir, H. (2025). Bonding Techniques–Interconnects. In Introduction to Microelectronics Advanced Packaging Assurance (pp. 21-39). Cham: Springer Nature Switzerland.

[29] Arab, Ali, and Qianmei Feng. "Reliability research on micro-and nano-electromechanical systems: a review." The International Journal of Advanced Manufacturing Technology 74.9 (2014): 1679-1690.

[30] Bhushan, Bharat. "Nanotribology and nanomechanics of MEMS/NEMS and BioMEMS/BioNEMS materials and devices." Microelectronic Engineering 84.3 (2007): 387-412.

[31] Pieters, Philip. "Wafer level packaging of micro/nanosystems." 5th IEEE Conference on Nanotechnology, 2005.. IEEE, 2005.

[32] S. Melle et al., "Reliability Overview of RF MEMS Devices and Circuits," 2003 33rd European Microwave Conference, Munich, Germany, 2003, pp. 37-40, doi: 10.1109/EUMA.2003.340820.

[33] Y. C. Lee, B. Amir Parviz, J. A. Chiou and Shaochen Chen, "Packaging for microelectromechanical and nanoelectromechanical systems," in IEEE Transactions on Advanced Packaging, vol. 26, no. 3, pp. 217-226, Aug. 2003, doi: 10.1109/TADVP.2003.817973.